# Non-paraxial design and fabrication of a compact OAM sorter in the telecom infrared

G. Ruffato,[1,2] M. Massari,[2,3] M. Girardi,[1,2] G. Parisi,[4] M. Zontini,[4] and F. Romanato[1,2,3,4*]

[1]Department of Physics and Astronomy 'G. Galilei', University of Padova, via Marzolo 8, 35131 Padova, Italy
[2]LANN, Laboratory for Nanofabrication of Nanodevices, EcamRicert, Corso Stati Uniti 4, 35127 Padova, Italy
[3]CNR-INFM TASC IOM National Laboratory, S.S. 14 Km 163.5, 34149 Basovizza, Trieste, Italy
[4]SM Optics – SIAE Microelettronica Group, Via M. Buonarroti 21, 20093 Cologno Monzese, Milano, Italy
*filippo.romanato@unipd.it

**Abstract:** A novel optical device is designed and fabricated in order to overcome the limits of the traditional sorter based on *log-pol* optical transformation for the demultiplexing of optical beams carrying orbital angular momentum (OAM). The proposed configuration simplifies the alignment procedure and significantly improves the compactness and miniaturization level of the optical architecture. Since the device requires to operate beyond the paraxial approximation, a rigorous formulation of transformation optics in the non-paraxial regime has been developed and applied. The sample has been fabricated as 256-level phase-only diffractive optics with high-resolution electron-beam lithography, and tested for the demultiplexing of OAM beams at the telecom wavelength of 1310 nm. The designed sorter can find promising applications in next-generation optical platforms for mode-division multiplexing based on OAM modes both for free-space and multi-mode fiber transmission.

## 1. Introduction

Since the seminal paper of Allen and coworkers in 1992 [1], light beams carrying orbital angular momentum (OAM) have known increasing interest and applications in a wide range of fields [2]: particle trapping and tweezing [3], high-resolution microscopy [4, 5], astronomy [6], holography [7], information and communication technology, both in the classical [8] and single-photon regimes [9]. Such beams are endowed with a set of $\ell$ intertwined helical wavefronts wrapping around a central dark singularity, being $\ell$ the amount of OAM per photon in units of $\hbar$. While the polarization state is described in a two-dimensional basis, the OAM degree of freedom opens to an unbounded state space, in which light beams carrying different integer OAM values are orthogonal to each other [10]. The possibility to exploit many optical carriers propagating at the same frequency with no interference, has aroused particular attention in the telecom field [11], which is constantly pursued by the worldwide restless demand for increasing bandwidth [12]. Besides wavelength, polarization, time, and amplitude/phase, the modulation and multiplexing of OAM promise to provide a significant increase in the information capacity of today's optical networks [13], both for free-space [14] and optical fibers communication [15, 16]. However, the implementation of OAM-mode division multiplexing (OAM-MDM) in a real scenario is still restrained by the lack of the commercial devices to perform optically the typical operations required in an optical link.

In the specific, a pivotal part of a communication link based on OAM-MDM is represented by the so-called multiplexer, i.e. the device located at the transmitter side allowing to generate a bunch of collimated and superimposed orthogonal beams with different OAM values. At the receiver, the beams are sorted by the demultiplexer, usually constituted by a multiplexer working in reverse. In the last decades, many solutions have been designed and presented in literature, exhibiting different levels of efficiency, complexity, and integration [17].

One of the most effective and preferred methods is based on the exploitation of transformation optics implementing a *log-pol* coordinate change, as demonstrated for the first time by Padgett's group in 2010 [18]. The underlying idea consists in converting the azimuthal phase gradients typical of OAM beams into linear phase gradients, i.e. tilted plane waves, which can be focused at distinct points, i.e. demultiplexed, by means of a common lens. This transformation is performed optically by a sequence of two elements: the first one, the un-wrapper, is designed to transform the input annular distribution into a linear one, while the second, the phase-corrector, compensates for phase-distortions and restores the linear phase gradients. In its first realization, this sorting scheme was introduced using spatial light modulators (SLMs) [18], and later with bulky refractive optical components [19], providing higher efficiency. In the quest for miniaturization, refractive 3D micro-optics have been shown [20], however for limited extent of OAM range sorting. More recently, we took another significant step forward miniaturization and integration by realizing the optics in a diffractive form [21, 22]. The same setup has been demonstrated to work as a multiplexer, in reverse [22-24].

A critical limitation of this sorting architecture is the need for two distinct optical elements which are notoriously difficult to align, making it very arduous to obtain output beams of good quality unless the two elements are perfectly planar, coaxial and aligned to each other, especially with short focal lengths, i.e. high miniaturization level. This represents a critical limiting factor in view of a future industrial realization and packaging of the device, where micro-optics should altogether be perfectly aligned at the production stage. In a previous configuration [21], taking advantage of the doughnut intensity distribution of OAM modes, the unwrapper optics was centered around the phase-corrector, and a separated mirror was used for back-reflection. While that layout could greatly simplify the alignment operation, since the two optics resulted aligned by design, on the other hand it imposed severe limitations on the input intensity distribution, which could not exceed the annular area of the unwrapper, notwithstanding its larger dimension. Moreover, despite the reduction in the degrees of freedom, a second optical element, i.e. a mirror, was still required.

In this work, in order to further overcome those drawbacks, we conceive a new optical configuration where the two elements are arranged side-by-side on the same facet of a transparent quartz slab. By adding a tilt to the first element, i.e. the unwrapper, the unwrapping beam illuminates the sample area again after back-reflection, without overlapping with the first zone. However, under this conditions, the traditional formulation of the *log-pol* phase patterns is no longer valid, and a more rigorous approach is necessary in order to compute the new phase patterns in the non-paraxial regime. In addition, the back-side facet of the slab was made reflective, then integrating the mirror required for back-reflection into a single optical device. A sample has been fabricated and designed for the telecom wavelength of 1310 nm on a commercial quartz slab, and tested for the sorting of OAM beams in the range from $\ell=-10$ to $\ell=+10$.

This new architecture addresses the important issue of being miniaturized down to few millimeters size and compatible with mass-production techniques. Moreover, the dramatic reduction in the degrees of freedom can significantly improve the integration level and provide a compact and full-optical device for next-generation optical boards implementing OAM-mode division multiplexing.

## 2. Theory and design

The demultiplexing method based on transformation optics performs a conformal mapping between orbital angular momentum and linear momentum states. The unitary transformation is achieved by a sequence of two elements [18]: un-wrapper and phase-corrector. The first one implements the desired *log-pol* coordinate transformation [26, 27] between two points (*x, y*) and (*u, v*), where $u=-a\cdot\log(r/b)$ and $v=a\cdot\varphi$, being (*r, φ*) polar coordinates on the plane *xy*, and (*a, b*) design parameters The second element restores the linear phase gradient compensating for the different paths travelled by each point of the wavefront. In other words, it performs the inverse transformation when the sequence is illuminated in reverse. Finally, a Fourier lens focuses the beam at a position $y_\ell=\ell\cdot\Delta s$, proportional to the OAM value, being $\Delta s=f\lambda/(2\pi a)$, as depicted in Fig. 1(a).

(a) *Standard configuration:*

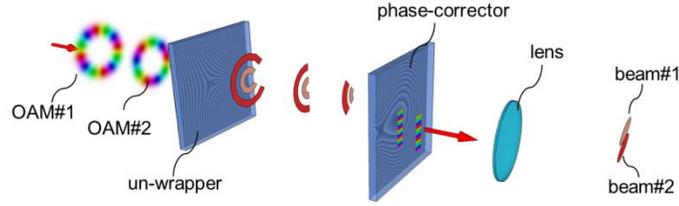

(b) *Non-paraxial compact configuration:*

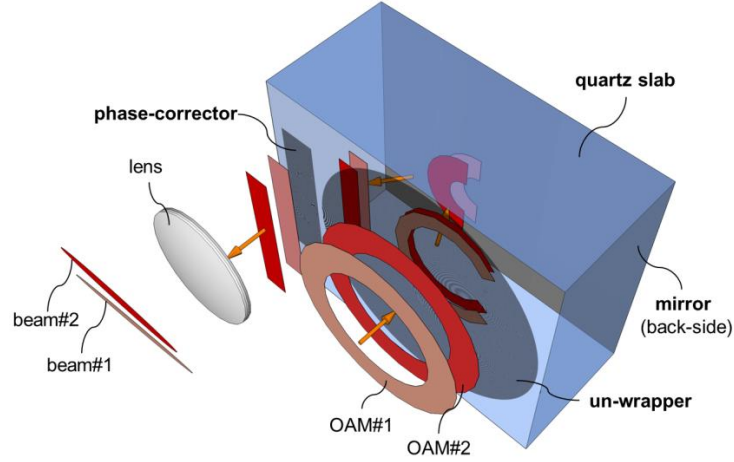

Fig. 1. (a) Scheme of the traditional *log-pol* sorter working principle with separated and coaxial optical elements, i.e. un-wrapper and phase-corrector. The azimuthal phase gradients of the input OAM-beams are transformed into linear phase gradients and focused at distinct positions at the focal plane of a lens. (b) Non-paraxial compact configuration with the two elements fabricated on the same surface of a single transparent slab with a reflective back-side. By adding a tilt to the first phase pattern, i.e. the un-wrapper, the unwrapping beam propagates with a non-null angle and after back-reflection illuminates the second element performing phase-correction. The two elements are patterned side-by-side on the same facet of the optical device.

In the new configuration, the transformed beam needs to be axially-displaced by a quantity *c*, therefore the optical transformation is defined by:

$$\begin{cases} u = -a\cdot\ln\left(\dfrac{\sqrt{x^2+y^2}}{b}\right)+c \\ v = a\cdot\arctan\left(\dfrac{y}{x}\right) \end{cases}. \quad (1)$$

With the introduction of the shift term *c*, the phase pattern of the un-wrapper in the paraxial approximation is given by:

$$\Omega_{UW}^{(p)} = \frac{2\pi a}{\lambda f}\left[y\cdot\arctan\left(\frac{y}{x}\right) - x\cdot\log\left(\frac{\sqrt{x^2+y^2}}{b}\right) + x\right] - \frac{2\pi}{\lambda}\frac{x^2+y^2}{2f} + \frac{2\pi c}{\lambda f}x, \qquad (2)$$

being $f$ the distance between un-wrapper and phase-corrector. For $c=0$, the traditional expression is obtained [18]. As depicted in Fig. 1(b), the unwrapping beam is back-reflected and illuminates the second pattern, i.e. the phase-corrector, placed side-by-side on the same plane with respect to the un-wrapper. It is worth noting that the beam propagation is inside a dielectric medium, i.e. no longer in air, therefore a precise estimation of the refractive index for the selected working wavelength is needed in order to optimize the design of the optical path. In addition, the focal length of the two confocal elements must be twice the thickness of the transparent slab.

Assuming the paraxial formulation in Eq. (2) for the un-wrapper term, we performed a numerical simulation of the output beam for increasing values of the off-axis shift $c$. As Fig. 2 shows for an input Laguerre-Gaussian (LG) beam with OAM $\ell=+5$, the deviation from the linear intensity pattern increases with the lateral displacement of the beam. The distortion dramatically affects the final spot shape, and therefore the capability of the sorter to correctly separate OAM beams.

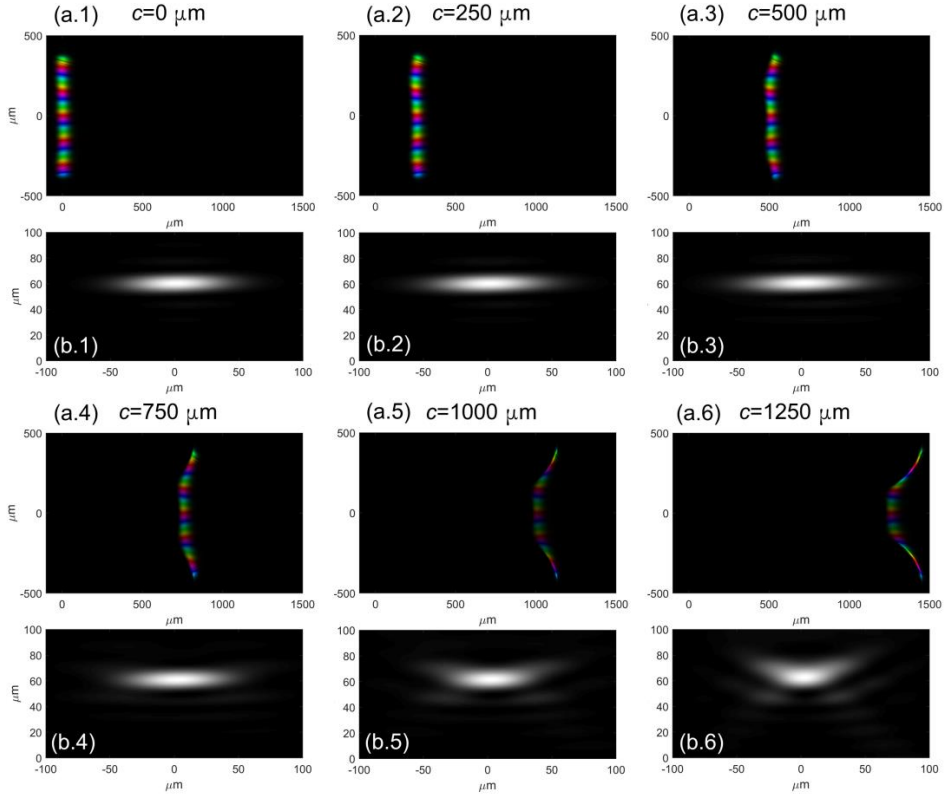

Fig. 2. (a) Simulation of the transformed beam for input LG mode with $\ell=+5$. Phase-corrected output beam assuming the traditional formulation of the un-wrapper in the paraxial approximation (Eq. (2)) for increasing values of the axial displacement $c$, in the range from 0 to 1250 μm, step 250 μm (a.1-6), and corresponding far-field spots (b.1-6). Parameters of the *log-pol* transformation: $a=120$ μm, $b=700$ μm, $f=4.572$ mm. Working wavelength $\lambda=1310$ nm. Refractive index of the medium $n=1.4467$. In (a.1-6) brightness and colors refer to intensity and phase, respectively.

In order to calculate the unwrapper phase pattern in the non-paraxial regime, we apply the stationary phase approximation to the Rayleigh-Sommerfeld integral. The field at a distance $f$ for an input plane-wave $U^{(in)}$ illuminating the phase-only un-wrapper with phase function $\Omega_{UW}$ is given by [25]:

$$U(u,v) = \frac{f}{i\lambda} \iint U^{(in)}(x,y) e^{i\Omega_{UW}(x,y)} \frac{e^{ik\sqrt{f^2+(x-u)^2+(y-v)^2}}}{f^2+(x-u)^2+(y-v)^2} dxdy . \qquad (3)$$

By applying the stationary phase approximation [27], we assume that all contributions from the above integral vanish except about the saddle points of the phase function:

$$\Phi(x,y) = \Omega_{UW}(x,y) + k\sqrt{f^2+(x-u)^2+(y-v)^2} . \qquad (4)$$

The condition $\nabla\Phi = 0$ leads to the following system of partial derivatives of $\Omega_{UW}$ unknown:

$$\begin{cases} \dfrac{\partial \Omega_{UW}}{\partial x} = -k \dfrac{x-u}{\sqrt{f^2+(x-u)^2+(y-v)^2}} \\ \dfrac{\partial \Omega_{UW}}{\partial y} = -k \dfrac{y-v}{\sqrt{f^2+(x-u)^2+(y-v)^2}} \end{cases} . \qquad (5)$$

Substituting the relations for the coordinates $(u, v)$ as defined by the *log-pol* optical transformation in Eq. (1), the un-wrapper phase function can be obtained from a numerical integration. Afterwards, the phase pattern $\Omega_{PC}$ of the corresponding phase-corrector can be calculated by numerically propagating a Gaussian plane wave illuminating the un-wrapper term. The rigorous solution of the diffracted field $U$ can be expressed in the convolution form of Eq. (3) as:

$$U(u,v) = FT^{-1}\left\{ FT\left\{ U^{(in)} e^{i\Omega_{UW}(x,y)} \right\} FT\left\{ \frac{f}{i\lambda} \frac{e^{ik\sqrt{f^2+x^2+y^2}}}{f^2+x^2+y^2} \right\} \right\} , \qquad (6)$$

where $FT$ and $FT^{-1}$ are the Fourier transform and its inverse, respectively. Finally, the phase profile for the phase-corrector is given by:

$$\Omega_{PC}(u,v) = 2\pi - \arctan\left( \mathrm{Im}\{U(u,v)\} / \mathrm{Re}\{U(u,v)\} \right). \qquad (7)$$

As Fig. 3 shows, this new formulation of the un-wrapper term (and of the phase-corrector, consequently), allows performing the correct coordinate transformation. The input beam is correctly unwrapped and illuminates the phase-corrector plane with a linear intensity distribution. The beam emerging from the phase-corrector presents a linear phase gradient, without the intensity and phase distortions introduced by a paraxial formulation of the unwrapper (as in Fig. 2(a.6) and 2(b.6)).

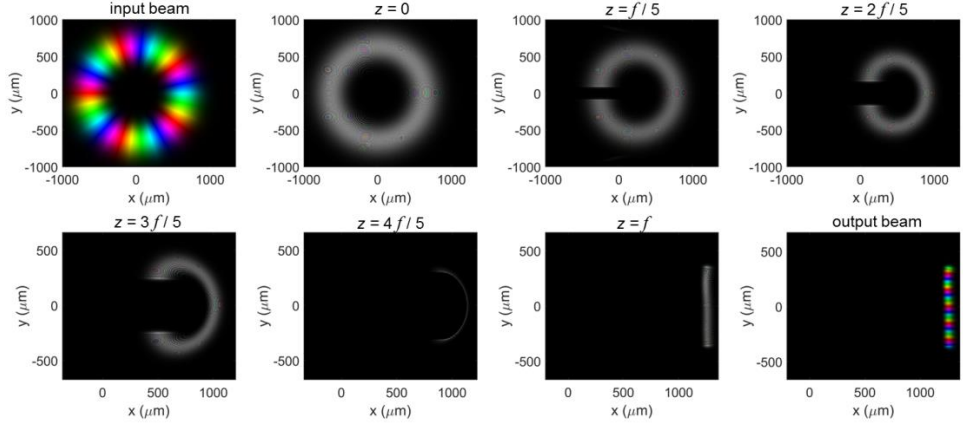

Fig. 3. Numerical simulation of a LG beam with $\ell=+5$ after illuminating the un-wrapper element, placed in $z=0$, calculated in the non-paraxial regime. With respect to Fig. 2(a.6), the beam is correctly transformed at the focal plane ($z=f$), and the linear phase gradient is retained after illuminating the corresponding phase-corrector. Parameters of the *log-pol* transformation: $a=120$ μm, $b=700$ μm, $f=4.572$ mm, $c=1.250$ mm. Working wavelength $\lambda=1310$ nm. Refractive index of the medium $n=1.4467$. Brightness and colors refer to intensity and phase, respectively.

## 3. Fabrication

The designed phase-patterns have been fabricated as surface-relief diffractive optics using high-resolution electron-beam lithography (EBL) on a resist layer [28, 29]. By locally controlling the released electronic dose, a different dissolution rate is induced in each zone of the exposed polymer, giving rise to a spatially-variant thickness after the development process. A dose-depth correlation curve, the so-called contrast curve, coupled with an accurate proximity compensation of the local dose, was required to calibrate the correct electron-dose and obtain the desired thickness for the surface relief pattern.

In this work, the optics have been fabricated by patterning a layer of negative resist (AR-N 7720.30, Allresist), spin-coated on a quartz slab (resist thickness around 3 μm), and pre-baked for 30 min at 85°C. The thickness of the slab is 2.286 mm, and the refractive index of the medium, measured with spectroscopic ellipsometry (J.A. Woollam VASE, 0.3 nm spectral resolution, 0.005° angular resolution) is 1.4467 at the working wavelength of 1310 nm. A 100-nm thick chromium layer was evaporated at the back side of the quartz slab. The resist exposure was performed with a JBX-6300FS JEOL EBL machine in high-resolution mode, 12 MHz, generating at 100 KeV and 100 pA an electron-beam with a diameter of 2 nm, providing a resolution down to 5 nm. At the experimental wavelength of the laser ($\lambda = 1310$ nm), the refractive index of the resist was assessed to be $n_R = 1.6450$, as measured again with the spectroscopic ellipsometer. For a given phase pattern of $\Omega(x, y)$, the depth $t(x, y)$ of the exposed zone for normal incidence in air is given by:

$$t(x, y) = \frac{\lambda}{n_R(\lambda)-1} \cdot \frac{2\pi - \Omega(x, y)}{2\pi}. \qquad (8)$$

The fabricated optics were made of square-pixels matrices with 256 phase levels. For the given wavelength, the maximal depth of the surface relief pattern was found to be 2031 nm using Eq. (8), with a thickness resolution of $\Delta t = 8$ nm in the case of 256 steps. By using custom numerical codes, the phase pattern of the simulated optics was converted into a 3D multilevel structure, which was in turn transformed into a map of electronic doses. A dose correction for compensating the proximity effects was required, in order both to match the layout depth with the fabricated relief and to obtain a good shape definition, especially in correspondence of phase discontinuities.

After exposure, a crosslinking bake was performed at 100°C for 60 minutes, followed by a post-baking process at 70°C for 2 hours in order to improve surface roughness. Finally, samples were developed for 510 s in AR 300-47 developer (Allresist). After development, the optical elements were gently rinsed in deionized water and blow-dried under nitrogen flux. The quality of the fabricated diffractive optics was inspected with scanning-electron microscopy (SEM), as shown in Fig. 4.

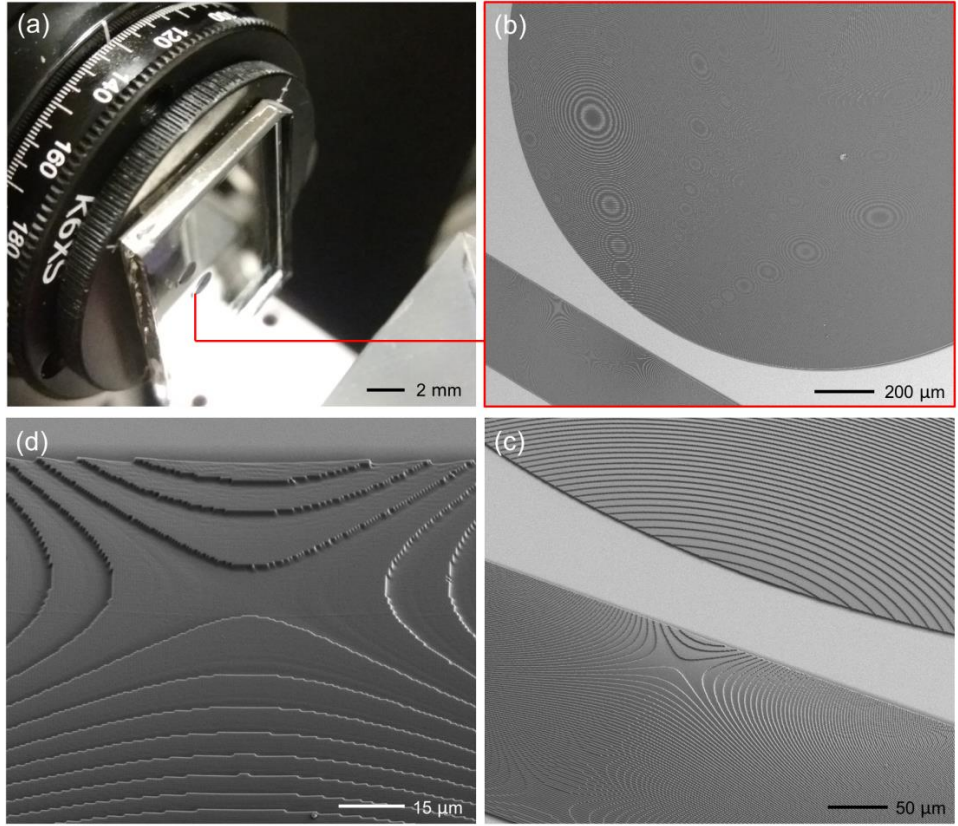

Fig. 4. (a) Picture of the fabricated sorter mounted on the sample holder of the experimental setup in Fig. 5. (b) SEM inspection of the zone between un-wrapper and phase-corrector and phase-corrector details at higher magnifications (c, d).

## 4. Optical characterization

The optical behavior of the sample has been tested for illumination under Laguerre-Gaussian (LG) beams carrying orbital angular momentum, generated with a liquid-crystal on Silicon (LCoS) spatial light modulator (X13267-08, Hamamatsu, pixel pitch 12.5 μm) using a phase and amplitude modulation technique [30]. The output of a DFB laser ($\lambda$=1310 nm) was collimated at the end of the single mode fiber with an aspheric lens of focal length $f_F$=7.5 mm (A375TM-C), linearly polarized and expanded with a first telescope ($f_1$=3.5 cm, $f_2$=10.0 cm) before illuminating the display of the SLM. A 50:50 beam-splitter was inserted after the telescope.

Then, a 4-$f$ system ($f_3$=20.0 cm, $f_4$=12.5 cm) with an aperture in the Fourier plane was used to isolate and image the first-order encoded mode on the sorter. A second 50:50 beam-splitter was used to split the beam and check the intensity profile with a first camera (WiDy

SWIR 640U-S, pixel pitch 15 µm). Afterwards, the OAM beam illuminated the first patterned zone of the device, mounted on a 6-axis kinematic mount (K6XS, Thorlabs). The planarity of the sample was correctly tuned in order to impinge normally on the un-wrapper and make the back-reflected transformed-beam illuminate the phase-correcting pattern properly. Then, the far-field of the beam emerging from the front-side facet was collected by a second camera (WiDy SWIR 640U-S) placed at the back-focal plane of a lens with focal length $f_5$=7.5 cm.

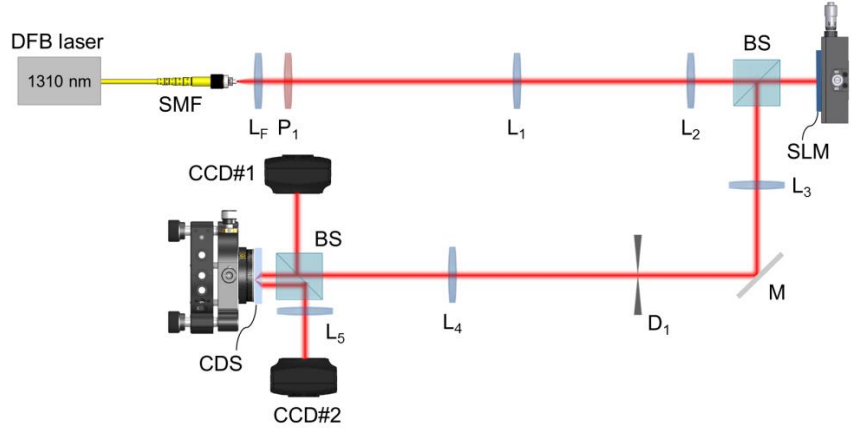

Fig. 5. Scheme of the experimental setup used for the optical characterization of the fabricated sorter. The DFB laser output ($\lambda$=1310 nm) is collimated at the end of the single mode fiber with an aspheric lens with focal length $f_F$=7.5 mm, linearly polarized ($P_1$) and expanded ($f_1$=3.5 cm, $f_2$=10.0 cm). The SLM first order is filtered ($D_1$) and resized ($f_3$=20.0 cm, $f_4$=12.5 cm) before illuminating the sorter. A beam splitter (BS) is used both to check the input beam with a first camera (CCD#1) and collect the sorter output at the focal plane of a fifth Fourier lens ($f_5$=7.5 cm) with a second camera (CCD#2). The compact diffractive sorter (CDS) is mounted on a 6-axis sample holder.

The optical performance was characterized by illuminating the sorter with LG beams carrying OAM in the range from -10 to +10 and recording the output intensity profiles. In Fig. 6, the input and output intensity patterns are reported for illumination under superposition of two LG beams with opposite OAM value with $\ell$ from 0 to 10. The superposition of two LG beams with opposite $\ell$ presents an intensity distribution with $2\ell$ petals, as shown in Fig. 6(a). As expected, the sorter separates correctly the two contributions, giving rise to two distinct spots whose axial displacement increases linearly with the OAM value (Fig. 6(b)). In Fig. 7, the position of the far-filed spot is plotted as a function of the input $\ell$, exhibiting a perfect accordance with the theory.

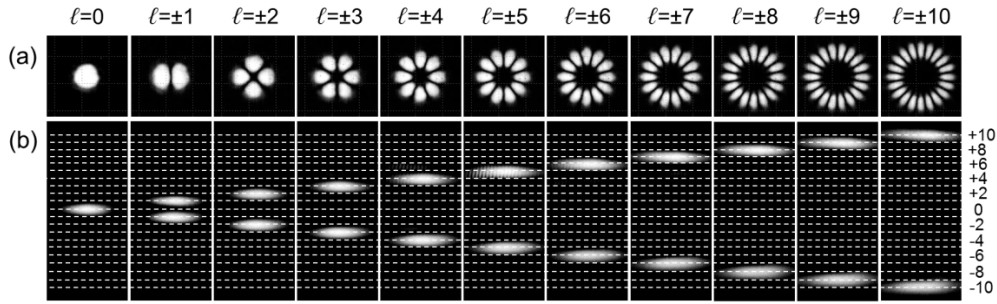

Fig. 6. Experimental data for input superposition of LG beams with opposite values of OAM (a) and corresponding output of the fabricated sorter (b). For each $\ell$ value increasing in modulus from 0 to 10, the superposition of two LG beams with opposite OAM generates a petalized beam with $2\ell$ petals. The two contributions are correctly detected and demultiplexed by the sorter at two distinct positions, proportionally to the value of $\ell$.

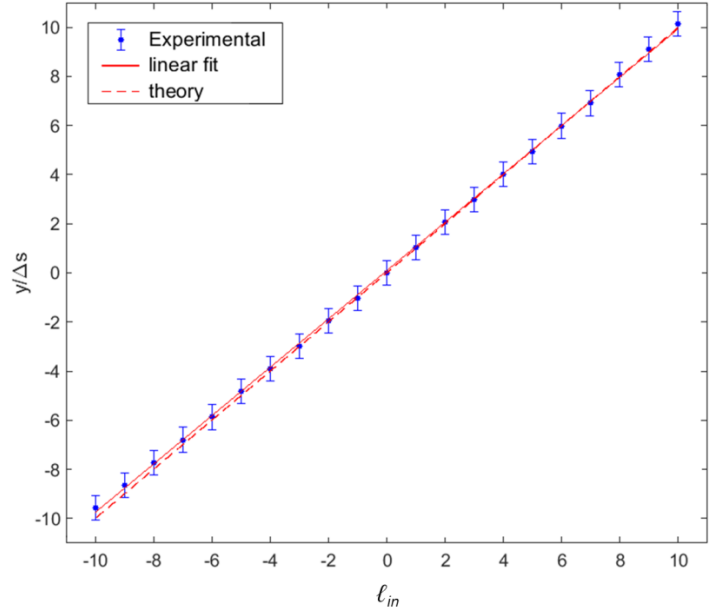

Fig. 7. Positions of the far-field spots for input LG beams carrying OAM $\ell_{in}$ in the range from -10 to +10. Experimental data (blue dots), linear fit (solid red line) and theoretical trend (dashed red line). The coordinate has been normalized by the parameter $\Delta s = \lambda f/2\pi a$, therefore the theoretical slope equals +1. Experimental slope: $+0.98 \pm 0.02$.

The area of the camera was divided into rectangular regions of interest, centered on each elongated spot in far-field and with the width given by the minimum distance between any two adjacent channels. By integrating the total intensity, the relative modal power and modal cross-talk $XT_j$ of the $j$th OAM channel could be determined using the definition [29]:

$$XT_j = 10 \cdot \log_{10}\left(\frac{I_{j,ALL/j}}{I_{j,ALL}}\right), \qquad (9)$$

being $I_{j,ALL}$ the signal at the $j$th channel when all the input channels were on, $j$ included, while $I_{ALL/j}$ is the same measure when the $j$th input channel was off.

As shown in Fig. 8, the system exhibits a good diagonal response, but the slight overlap between adjacent channels, due to the non-null width of the far-field spots (see Fig. 6(b)) causes a spread of the input energy over the first-neighboring OAM values, which badly affects the inter-channel cross-talk. The average cross-talk resulted to be $-(5.28 \pm 1.21)$ dB.

A further improvement could be achieved by considering a sparse mode set, i.e. choosing non-consecutive OAM values, as shown in [21], however at the expense of discarding many channels. An alternative solution consists in including a fan-out element in cascade [31] to create multiple copies extending the phase gradient of the sorted beam, which is focused at the same position as before but with a narrower width, thus improving the separation between spots. This solution can dramatically reduce the channel overlap without sacrificing modal density. The fan-out optical operation could be included in the un-wrapper element, and the phase-corrector redesigned accordingly, as shown in [32, 33], therefore maintaining the compactness and integration level of the optical device.

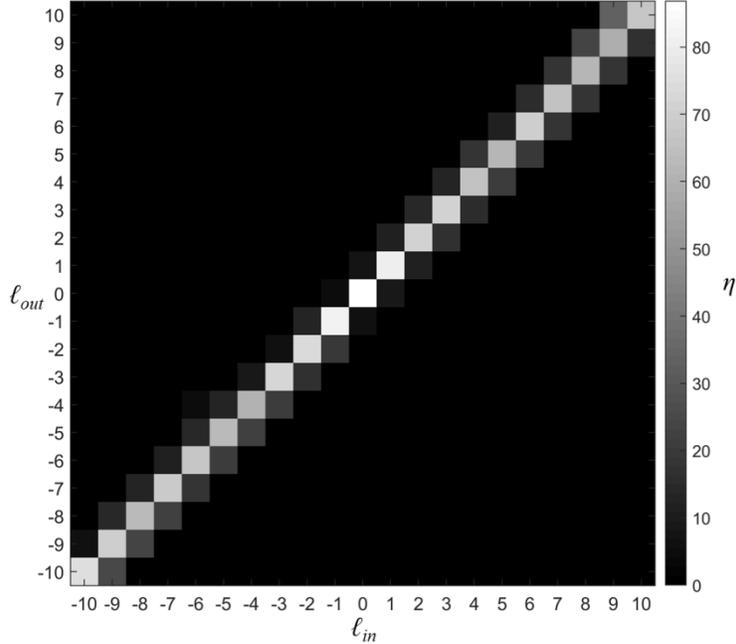

Fig. 8. Efficiency map for input OAM beams with input OAM value $\ell_{in}$ in the range from $\ell = -10$ to $\ell = +10$.

## 5. Conclusions

A formulation of *log-pol* transformation optics has been developed in the non-paraxial regime, and exploited for the design of a novel compact device for OAM demultiplexing. The new approach replaces the usual phase patterns of the traditional sorter, whose validity was limited to the paraxial approximation, and integrates all the required optical elements into a single compact platform.

The two required phase-patterns, i.e. un-wrapper and phase-corrector, have been fabricated in the form of phase-only diffractive optics by patterning a thin layer of resist with high-resolution electron-beam lithography. By adding an axial displacement to the optical transformation, the two patterns have been arranged side-by-side on the same substrate, and a reflective surface has been used for back-reflecting the unwrapping beam onto the phase-correcting pattern. In the proposed configuration, the diffractive patterns and the reflective surface belong to two opposite sides of a single transparent slab, dramatically reducing the degrees of freedom and significantly improving compactness, integration and alignment. As a matter of fact, the fabrication technique and the new optical scheme remarkably improve the miniaturization level and the compactness of the system, making feasible its integration. This solution avoids bulky refractive/reflective elements and provides a compact single-block sorter, without the need for aligning many separated elements. The optical device has been designed for the telecom wavelength of 1310 nm, and optically characterized for the demultiplexing of beams with OAM in the range from -10 to +10. The possibility to replicate the fabricated optics with fast mass-production techniques, such as nano-imprint lithography [33, 34], which can provide high-throughput and much lower production costs, makes these optical elements promising for integration into optical platforms for MDM.

These results pave the way for practical OAM multiplexing and demultiplexing devices for use both in classical and quantum communication. In addition, the approach underlying the design of transformation optics in the non-paraxial regime, with the application of the stationary phase approximation to the Rayleigh-Sommerfeld propagation integral, could be extended and applied to the design of a wide range of optical elements working beyond the paraxial approximation.

**Funding**

This research was funded by CEPOLISPE (project VORTEX 2), by SM Optics S.r.l.–SIAE Group, and by the University of Padova (project LIFE LAB).

**Disclosures**

The authors declare that there are no conflicts of interest related to this article.

**References**

1. L. Allen, M.W. Beijersbergen, R.J.C. Spreeuw, and J.P. Woerdman, "Orbital angular momentum of light and the transformation of Laguerre-Gaussian modes," Phys. Rev. A **45**, 8185-8189 (1992).
2. M.J. Padgett, "Orbital angular momentum 25 years on," Opt. Express **25**, 11265-11274 (2017).
3. M.J. Padgett, and R. Bowman, "Tweezers with a twist," Nature Photon. **5**, 343–348 (2011).
4. M. Ritsch-Marte, "Orbital angular momentum light in microscopy," Phil. Trans. R. Soc. A **375**, 20150437 (2017).
5. G. Vicidomini, P. Bianchini, and A. Diaspro, "STED super-resolved microscopy," Nat. Methods **15**, 173-182 (2018).
6. E. Mari, F. Tamburini, G.A. Swartzlander, A. Bianchini, C. Barbieri, F. Romanato, and B. Thidé, B, "Sub-Rayleigh optical vortex coronagraphy," Opt. Express **20**, 2445-2451 (2012).
7. G. Ruffato, R. Rossi, M. Massari, E. Mafakheri, P. Capaldo, and F. Romanato, "Design, fabrication and characterization of Computer-Generated Holograms for anti-counterfeiting applications using OAM beams as light decoders," Sci. Rep. **7**, 18011 (2017).
8. J. Wang, "Twisted optical communications using orbital angular momentum," China Phys. Mech. Astron. **62**, 34201 (2019).
9. M. Mirhosseini, O.S. Magana-Loaiza, M.N. O'Sullivan, B. Rudenburg, M. Malik, M.P.J. Lavery, M.J. Padgett, M.J., D.J. Gauthier, and R.W. Boyd, "High-dimensional quantum cryptography with twisted light," New J. Phys. **17**, 033033-1-12 (2015).
10. D. Andrews, and M. Babiker, *The angular momentum of light* (Cambridge University Press, 2013).
11. E. Agrell, M. Karlsson, A. R. Chraplyvy, D. J. Richardson, P. M. Krummrich, P. Winzer, K. Roberts, J. K. Fischer, S. J. Savory, B. J. Eggleton, M. Secondini, F. R. Kschischang, A. Lord, J. Prat, I. Tomkos, J. E. Bowers, S. Srinivasan, M. Brandt-Pearce, and N. Gisin, "Roadmap of optical communications," J. Opt. **18**(6), 063002 (2016).
12. P. J. Winzer, D. T. Neilson, and A. R. Chraplyvy, "Fiber-optic transmission and networking: the previous 20 and the next 20 years," Opt. Express **26**(18), 24190–24239 (2018).
13. S. Yu, "Potentials and challenges of using orbital angular momentum communications in optical interconnects," Opt. Express **23**(3), 3075–3087 (2015).
14. A.E. Willner, Y. Ren, G. Xie, Y. Yan, L. Li, Z. Zhao, J. Wang, M. Tur, A.F. Molish, and S. Ashrafi, "Recent advances in high-capacity free-space optical and radio-frequency communications using orbital angular momentum multiplexing," Philos Trans A Math Phys Eng Sci. **375**, 20150439-1-12 (2017).
15. S. Ramachandran, and P. Kristensen, "Optical vortices in fiber," Nanoph. **2**, 455–474 (2013).
16. N. Bozinovic, Y. Yue, Y. Ren, N. Tur, P. Kristensen, H. Huang, A.E. Willner, and S. Ramachandran, "Terabit-scale orbital angular momentum mode division multiplexing in fibers," Science **340**, 1545-1548 (2013).
17. C. Wan, G. Rui, J. Chen, and Q. Zhan, "Detection of photonic orbital angular momentum with micro- and nano-optical structures," Front. Optoelectron. **1**(12), 88-96 (2019).
18. G. C. G. Berkhout, M. P. J. Lavery, J. Courtial, M. W. Beijersbergen, and M. J. Padgett, "Efficient sorting of orbital angular momentum states of light," Phys. Rev. Lett. **105**(15), 153601 (2010)
19. M. P. J. Lavery, D. J. Robertson, G. C. G. Berkhout, G. D. Love, M. J. Padgett, and J. Courtial, "Refractive elements for the measurement of the orbital angular momentum of a single photon," Opt. Express **20**(3), 2110–2115 (2012).
20. S. Lightman, G. Hurvitz, R. Gvishi, and A. Arie, "Miniature wide-spectrum mode sorter for vortex beams produced by 3D laser printing," Optica **4**, 605-610 (2017).
21. G. Ruffato, M. Massari, and F. Romanato, "Compact sorting of optical vortices by means of diffractive transformation optics," Opt. Lett. **42**(3), 551–554 (2017).
22. G. Ruffato, M. Massari, G. Parisi, and F. Romanato, "Test of mode-division multiplexing and demultiplexing in free-space with diffractive transformation optics," Opt. Express **25**(7), 7859–7868 (2017).
23. R. Fickler, R. Lapkiewicz, M. Huber, M.P.J. Lavery, M.J., Padgett, and A. Zeilinger, "Interface between path and orbital angular momentum entanglement for high-dimensional photonic quantum information," Nat. Comm. **5**, 4502 (2014).

24. W. Li, K.S. Morgan, Y. Li, K. Miller, G. White, R.J. Watkins, and E.G. Johnson, "Rapidly tunable orbital angular momentum (OAM) system for higher order Bessel beams integrated in time (HOBBIT)," Opt. Express **27**, 3920-3934 (2019).
25. J. Li, Z. Peng, and Y. Fu, "Diffraction transfer function and its calculation of classic diffraction formula," Opt. Comm. **280**, 243-248 (2007).
26. Y. Saito, S. Komatsu, and H. Ohzu, "Scale and rotation invariant real time optical correlator using computer generated hologram," Opt. Commun. **47**(1), 8–11 (1983).
27. W. J. Hossack, A. M. Darling, and A. Dahdouh, "Coordinate transformations with multiple computer-generated optical elements," J. Mod. Opt. **34**(9), 1235–1250 (1987).
28. M. Massari, G. Ruffato, M. Gintoli, F. Ricci, and F. Romanato, "Fabrication and characterization of high-quality spiral phase plates for optical applications," Appl. Opt. **54**(13), 4077–4083 (2015).
29. G. Ruffato, M. Massari, and F. Romanato, "Diffractive optics for combined spatial- and mode- division demultiplexing of optical vortices: design, fabrication and optical characterization," Sci. Rep. **6**, 24760-1-12 (2016).
30. C. Rosales-Guzmán and A. Forbes, *How to Shape Light With Spatial Light Modulators* (SPIE Press, 2017).
31. M. Mirhosseini, M. Malik, Z. Shi, and R. W. Boyd, "Efficient separation of the orbital angular momentum eigenstates of light," Nat. Commun. **4**(1), 2781 (2013).
32. C. Wan, J. Chen, and Q. Zhan, "Compact and high-resolution optical orbital angular momentum sorter," APL Photonics **2**(3), 031302 (2017).
33. G. Ruffato, M. Girardi, M. Massari, E. Mafakheri, B. Sephton, P. Capaldo, A. Forbes, and F. Romanato, "A compact diffractive sorter for high-resolution demultiplexing of orbital angular momentum beams," Sci. Rep. **8**(1), 10248 (2018).
34. B. J. Wiley, D. Qin, and Y. Xia, "Nanofabrication at high throughput and low cost," ACS Nano **4**(7), 3554–3559 (2010).